\documentclass[aps,prl,reprint,superscriptaddress,showpacs]{revtex4-1}

\usepackage{amssymb}
\usepackage{graphicx}
\usepackage{bm,url}

\begin{document}

\title{Liquid is More Rigid than Solid in a High-Frequency Region}

\author{Naoki Hasegawa}
\affiliation{Department of Basic Science, 
University of Tokyo, 3-8-1 Komaba, Meguro, 
Tokyo 153-8902, Japan}
\author{Tatsuro Yuge}
\email[]{yuge.tatsuro@shizuoka.ac.jp}
\affiliation{Department of Physics, Shizuoka University, 836 Ohya Suruga Ward, Shizuoka 422-8529, Japan}
\author{Akira Shimizu}
\email[]{shmz@ASone.c.u-tokyo.ac.jp}
\homepage[]{http://as2.c.u-tokyo.ac.jp}
\affiliation{Department of Basic Science, 
University of Tokyo, 3-8-1 Komaba, Meguro, 
Tokyo 153-8902, Japan}

\date{August 14, 2015}

\begin{abstract}
We compare rigidity of materials in two phases, liquid and solid phases.
As a measure of the rigidity, 
we employ the one characterizing 
how firmly the material is fixed by low density of
pinning centers, such as impurities and rough surfaces of walls, 
against a weak force.
Although a solid is more rigid than a liquid against a low-frequency force, 
we find that against a high-frequency force 
the liquid becomes more rigid than the solid of the same material.
Since this result is derived from 
universal properties of a response function, 
it is valid for wide classes of materials, 
including 
quantum and classical systems and
crystalline and amorphous solids.
An instructive example is studied using 
nonequilibrium molecular dynamics simulations.
We find that the frequency region in which a 
solid is more flexible than a liquid is not purely determined 
by intrinsic properties of the solid.
It depends also on extrinsic factors such as the density of pinning centers.
\end{abstract}



\maketitle



Well-known relations sometimes have surprising implications,
which have not been realized for a long time.
%
We here present one, which concerns rigidity of materials.
The conventional wisdom says that a solid is more rigid than a liquid.
In a short time scale, however, it is known that some liquids become `rigid' 
(or `elastic') \cite{LL,materials,fluid,polymer,liq1}.
This raises fundamental questions:
Are solids `rigid' in such a time scale?
More deeply, 
{\em when comparing liquid and solid phases of the same material,
which is more `rigid' at high frequencies?}
The purpose of this paper is to give a universal answer to these questions.

The term `rigid' refers to 
the ability of a material to resist deformation by an applied force $\bm{f}$.
However, this does not define the rigidity uniquely,
because the ability 
depends strongly on the natures of $\bm{f}$,
such as its strength, frequency, and spatial variation 
\cite{materials}.
Furthermore, in general, two or more types of deformations
are induced, such as displacement and rotation \cite{materials}.
According to the types of deformation and the natures of $\bm{f}$,
the ability to resist the deformation is quantified by various measures,
such as the elastic modulus, stiffness and viscosity 
\cite{LL,materials,fluid,polymer,liq1}.
Therefore, we must first specify 
the type of deformation, the nature of $\bm{f}$, 
and the measure that defines the rigidity.
To study the above fundamental question,
we assume a simple case where 
$\bm{f}$ is sufficiently weak, so that 
neither hysteresis nor creep occurs
and the material responds to 
$\bm{f}$ linearly \cite{KTH,Zubarev,SY2010,Y2010,S2010,liq2,liq3}.
We also assume that $\bm{f}$ is uniform spatially but varies
as a function of time with frequency $\omega$.
Physically, $\bm{f}$ can be an electric field, 
gravitational or inertial force, and so on.
To probe rigidity of a material by 
applying such a simple force, 
we assume that microscopic `pinning centers' exist \cite{sliding1,sliding2}, such as impurities and roughness of walls of the container.
This leads us to a reasonable measure of rigidity.

We show that 
if a material is more rigid in a solid phase 
than in a liquid  phase 
at low frequencies then 
in a certain high-frequency region
{\em it is more rigid in the liquid phase than in the solid phase}.
Since this result is derived from 
universal properties of a response function
\cite{liq1,KTH,Zubarev,SY2010,Y2010,S2010,liq2,liq3},
it is a universal result
that holds for wide classes of materials,
including 
crystalline and amorphous solids.
We also give an instructive example, and explore physics.
We show that 
the frequency region in which a solid becomes more flexible than 
a liquid is not purely intrinsic to the material.
It depends also on extrinsic factors such as the impurity density.

{\em Measure of rigidity.---} 
As mentioned above, 
there are many measures of rigidity.
We here employ the following one.
We consider a material 
of size $L_x \times L_y \times L_z$,
which is subject to a weak force 
$\bm{f}(t) = (f(t), 0, 0)$. 
Particles composing the material are driven by $\bm{f}(t)$,
and 
a particle current is induced.
Since we are not interested in 
a particular cross section of the material, 
we consider the integral of the current over the position $x$ of the cross section.
Since the integral agrees with the total momentum divided by the 
particle mass,
we consider the momentum per particle, 
which we denote by $I(t)$, instead of the current.
For quantum systems, 
the operator for $I(t)$
is given by
\begin{equation}
\hat{I} =
{1 \over N_e} \sum_{j=1}^{N_e} \hat{p}^{j}_{x},
\label{eq:I}
\end{equation}
where
$N_e$ is the number of particles that respond to $\bm{f}(t)$,
$\hat{p}^{j}_{x}$ is the $x$ component of the momentum operator of the $j$-th particle. 
Under the aforementioned assumptions,
the Fourier components of $f(t)$ and $I(t)$ are related as
\begin{equation}
I_\omega = \Xi(\omega) f_\omega.
\end{equation}
Here, 
$
\Xi(\omega) = \Xi'(\omega) + i \, \Xi''(\omega)
$
is the complex admittance, 
which is a continuous function of $\omega$ \cite{KTH,Zubarev,SY2010}.
Its real part 
$\Xi'(\omega)$ 
determines the rate of dissipation,
whereas the imaginary part 
$\Xi''(\omega)$ is related to non-dissipative displacement \cite{KTH}.
%
%

We note that 
one can fix a rigid material firmly to a board   
by hammering a small number of nails into it,
whereas one can hardly fix a flexible material.
When one shakes the board, the inertial force acting on 
the material is time-dependent,
which generally causes dissipation in the material.
The dissipation would vanish if the material were a `rigid body' or an `elastic' 
one \cite{LL,materials}.
We now replace the nails with 
microscopic pinning centers, 
such as impurities, 
and the inertial force with a force $\bm{f}(t)$.
In the presence of pinning centers, a dissipative current does not flow 
in a solid phase 
at $\omega = 0$  (i.e., $\Xi'(0)=0$), 
whereas it does in a liquid phase (i.e., $\Xi'(0)>0$).
Considering also the continuity and the positivity of $\Xi'(\omega)$
(see below),
we see that 
{\em at low frequencies $\Xi'(\omega)$ is much smaller in a solid phase 
than in a liquid phase}.
As for the imaginary part $\Xi''(\omega)$,
by contrast, 
such a distinct tendency is absent (at least at low frequencies) 
because $\Xi''(0)=0$ {\em in all phases}.
Furthermore,
$\Xi''(\omega)$ can take negative values, 
and $\omega \Xi''(\omega)$ converges
to the {\em same} value as $\omega \to \infty$ \cite{KTH,Zubarev,SY2010}.

From these considerations, we say a material is 
{\em more rigid (flexible) at frequency $\omega$} 
if the real part $\Xi'(\omega)$ is smaller (larger)
in the presence of low density of pinning centers.
We will explicitly show later that according to this measure a material 
is more rigid at low frequencies in a solid phase than in a liquid phase, in consistency with the conventional wisdom.
Note that in Ref.~\cite{LL} 
`viscosity' (which corresponds to the flexibility here)
is quantified also by the magnitude of dissipation.
It is obvious that our measure of rigidity is also related 
to other measures at low $\omega$.
At high $\omega$, however, relations among different measures 
(including ours) are not well understood yet.

{\em General properties of admittance.---}
To get a {\em universal} answer to the aforementioned question, 
we note the following general properties of $\Xi'(\omega)$.
%

\noindent
(i) Continuity: 
$\Xi(\omega)$ is the Fourier transform of the response function $\Phi(t)$.
Since $I(t)$ should be finite even when $\bm{f}$ is constant independent of $t$, $\Phi(t)$ is integrable in general.
Hence, according to Theorem VI-1.2 of Ref.~\cite{FTmath}, 
{\em $\Xi(\omega)$ is uniformly continuous}.

\noindent
(ii) Positivity: 
According to the second law of thermodynamics, 
all materials in thermal equilibrium, such as crystalline solids, 
are {\em passive}, i.e., 
one cannot extract work from them.
Amorphous solids, such as glasses and amorphous semiconductors,
are out of equilibrium but macroscopic currents are absent 
(e.g., $I(t)=0$ when $\bm{f}(t)=0$).
They are also passive (to small $\bm{f}$), 
within the time scale where they are stable.
For all such passive stable materials 
in which macroscopic currents are absent,
%
the power supplied from $\bm{f}(t)$ is dissipated, 
and hence $\Xi'(\omega) \geq 0$ for all $\omega$.
%

\noindent
(iii) Sum rule:
We assume a typical case where $\bm{f}(t)$ interacts with 
the particles composing the material through the interaction 
Hamiltonian, 
\begin{equation}
- \sum_{j=1}^{N_e} \bm{f}(t) \cdot \hat{\bm{q}}^j
=
- \sum_{j=1}^{N_e} f(t)  \hat{q}^j_x. 
\label{eq:fq}
\end{equation}
Here, 
$\hat{\bm{q}}^j = (\hat{q}^j_x, \hat{q}^j_y, \hat{q}^j_z)$ 
is the position of the $j$-th particle.
In this case, the following sum rule holds {\em rigorously}
\cite{KTH,Zubarev,SY2010}:
\begin{equation}
\int_{-\infty}^{\infty} \Xi'(\omega) {d \omega \over \pi}
=
\Big\langle
{1 \over i \hbar} \Big[
\sum_{j=1}^{N_e} \hat{q}^j_x, \hat{I}
\Big]
\Big\rangle
=
1,
\label{sumrule}\end{equation}
where 
$\langle \ \rangle$ denotes 
the expectation value in the state before $\bm{f}$ is applied.
The right-hand side (RHS) is the $t=0$ value of the response function, whereas the left-hand side (LHS) decomposes it
into {\em all} time scales \cite{KTH,Zubarev,SY2010}.

Properties (i) and (ii) hold very generally, as explained above.
The sum rule (iii) also holds very generally, 
because it relies only on 
the forms of the observable of interest, Eq.~(\ref{eq:I}), 
and the interaction with $\bm{f}$, 
Eq.~(\ref{eq:fq}),
which are typical forms in both 
quantum and classical (nonrelativistic) mechanics.
It therefore holds rigorously irrespective of interactions among 
particles, even when pinning centers are present,
in both quantum and classical systems.
Although the conventional derivation of the sum rule 
\cite{liq1,KTH,Zubarev} assumed an equilibrium Gibbs state for 
$\langle \ \rangle$, 
we can show that it holds also for thermal pure quantum states 
\cite{SS2012,SS2013,HSSS2014}.
More importantly, it was recently proved to hold {\em rigorously} even for  
nonequilibrium (NE) states \cite{SY2010,Y2010,S2010}, including 
amorphous solids such as glasses and amorphous semiconductors.
Although phenomenological models sometimes fail to satisfy 
the sum rule, they should be modified at high frequencies in 
such a way that the sum rule is satisfied \cite{KTH}.

{\em Main result.---}
We note that the RHS of Eq.~(\ref{sumrule}) 
is independent of the state of the system, 
i.e., whether the system is in a liquid or solid phase.
Hence, although $\Xi'(\omega)$ at {\em each $\omega$}
changes drastically across a 
phase transition, 
its {\em sum} over all $\omega$ does {\em not} change at all.
Since $\Xi'(\omega)$ is continuous and positive from (i) and (ii) above,
this is a strong restriction. 
In particular, 
if a solid phase has smaller $\Xi'(\omega)$ 
in a low-$\omega$ region
(than a liquid phase, as expected)
then 
it must have larger $\Xi'(\omega)$ in a high-$\omega$ region.
We thus have the main result: 
{\em For all passive stable material in which macroscopic currents 
are absent (when driving forces are absent),
if it is more rigid in a solid phase at low frequencies, 
then it is more rigid in a liquid phase at certain high frequencies}.

This result is derived rigorously from the general properties (i)-(iii) 
of $\Xi'(\omega)$, which hold very generally as discussed above.
Therefore, the result also holds very generally,
 for both quantum and classical systems, 
for both crystalline (equilibrium) and amorphous (passive nonequilibrium) solids.
For glasses, for example, 
{\em a glass is more flexible in a `hard state' than in a `molten state' 
at certain high frequencies}.



{\em Example.---}
We have pointed out that the well-known properties (i)-(iii) of the admittance 
lead to a counterintuitive conclusion.
We now give an instructive example.
It also gives the answer to 
the following natural question:
%
%
Regarding the frequency region in which a 
solid is more flexible than a liquid,
is it an intrinsic property of the solid?
For each $\omega$ there would exist some phonon modes, 
which are determined by the crystal structure,
that are particularly sensitive to $\bm{f}(t)$. 
One might expect that 
such modes would determine the frequency region, and hence 
it  would be an intrinsic property of the solid.
According to this expectation, 
the frequency region would be 
insensitive to extrinsic factors such as 
the density and configuration of pinning centers.
We will show that this is {\em false}.
A single counter example suffices to show it.

As such an example we study a classical model using 
NE molecular dynamics (MD) simulations.
The model 
should undergo a liquid-solid phase transition.
Furthermore, 
it should have stable NE states in the presence 
of $\bm{f}(t)$, because otherwise $\Xi(\omega)$ could not be well-defined.
To meet these requirements, we generalize 
the models of Refs.~\cite{YIS2005,YS2007,YS_alder,YS2009, Lee},
which have stable NE states
and are free from problems due to the long-time tails
\cite{Alder1967,Ernst,Dorfman,Zwanzig,Kawasaki,Pomeau}.

\begin{figure}[htbp]
\begin{center}
\includegraphics[width=0.75\linewidth]{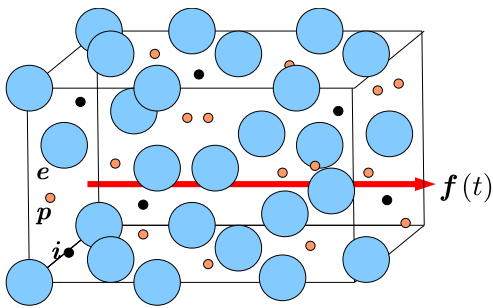}
\end{center}
\caption{(Color online)
Model for nonequilibrium MD simulations,
which consists of 
$N_e, N_p$ and $N_i$ $e,p$ and $i$ particles, respectively.
Periodic boundary conditions of size $L_x \times L_y \times L_z$
are imposed.
}
\label{fig1}
\end{figure}
As shown in Fig.~\ref{fig1}, 
our model consists of three types of particles, 
which are called $e,p$ and $i$
after a typical case where 
they are electrons, phonons and impurities, respectively.
Their radii are taken $1,0$ and $0$, respectively. 
For interactions between these particles,
we assume those close to the hard-core interactions 
\cite{interaction}.
Hence, 
with increasing the volume filling factor 
$\rho_e = (4 \pi/3)N_e/(L_x L_y L_z)$, 
$e$ particles 
undergo a first-order 
transition of the Alder type \cite{Alder1957,HR},
from a liquid phase ($\rho_e < \rho_e^{\rm lq}$) 
to a solid phase ($\rho_e > \rho_e^{\rm sl}$).
Both phases coexist for 
$\rho_e^{\rm lq} \leq \rho_e \leq \rho_e^{\rm sl}$.
For hard-sphere particles of unit radius, 
$\rho_e^{\rm lq} = 0.494$ and $\rho_e^{\rm sl}= 0.545$.
By contrast, 
neither $p$ nor $i$ particles undergo such a transition
because they have vanishing volume  filling factor 
(because of vanishing radii).
The $p$ particles are particles of the `environment,'
which is attached to a heat reservoir.
The $i$ particles 
are `impurities,' 
which are fixed at random positions,
and scatter $e$ 
particles via static potentials, working as pinning centers. 
A force $\bm{f}(t)$ 
acts only on $e$ particles.
Power absorbed by $e$ particles from $\bm{f}(t)$ is  
dissipated to the reservoir through the $e$-$p$ interaction.
As a result, a stable NE state is realized in 
the presence of $\bm{f}(t)$,
and the system has well-defined $\Xi(\omega)$.
Note that $p$ and $i$ particles are just auxiliary particles, 
and the model essentially describes a one-component open system of $e$ 
particles.
Note also that neither temperature 
nor mass $m_e$ is relevant to the phase transition of hard-sphere particles
because changing their values is equivalent to 
changing the time scale \cite{HR}. 
Hence we shall take $m_e=m_p=1$ and 
the temperature of the reservoir $T=1$.
We will present the results for $N_e=320, N_p=160$.
We have confirmed that no 
significant changes occur by increasing the system size \cite{spm}.
Details of the model and simulation are described in Ref.~\cite{spm}. 

\begin{figure}[htbp]
\begin{center}
\includegraphics[width=0.9\linewidth]{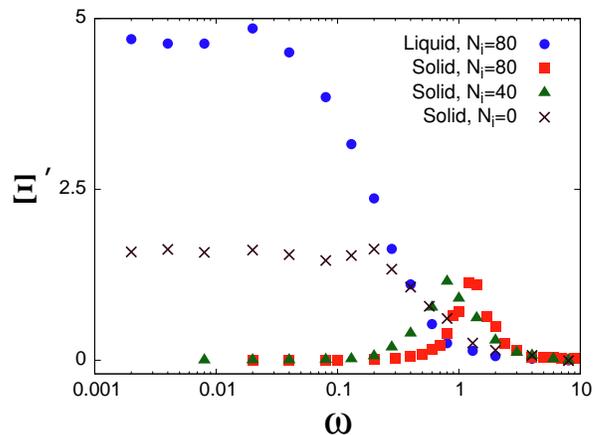}
\end{center}
\caption{(Color online)
Real part of $\Xi(\omega)$, 
for the liquid phase with  $(L_x, L_y, L_z)=(15,20,20)$ and $N_i=80$ (circles),
and for the solid phase with  $(L_x, L_y, L_z)=(15,12,12)$
and $N_i=80$ (squares), $N_i=40$ (triangles), and $N_i=0$ (crosses).
Note that the area below each curve does not give the sum value
because the horizontal axis is in a logarithmic scale.
}
\label{fig2}
\end{figure}

{\em Results.---}
We first study the liquid phase by taking 
$\rho_e = 0.223$ ($< \rho_e^{\rm lq} = 0.494$),
in the presence of impurities of
number density $n_i = N_i/(L_x L_y L_z) = 80/6000=0.0133\cdots$.
The circles in Fig.~\ref{fig2} show $\Xi'(\omega)$ in this case.
Its behavior is almost of the Drude type,
$\Xi'(\omega) \simeq \Xi'(0)/[1+(\omega \tau)^2]$, 
which is typical to liquid phases, 
where the least square fit gives the collision time $\tau \simeq 5$.
As expected, 
the liquid is flexible at low frequencies 
$\omega \lesssim 1/\tau \simeq 0.2$.
Numerical integration of $\Xi'(\omega)$ gives 
$\simeq 0.97$ 
for the LHS of Eq.~(\ref{sumrule}), 
in consistency with the sum rule
\cite{cutoff}.
This agreement evidences that 
our MD simulation well describes NE states driven by 
$\bm{f}(t)$ 
\cite{SY2010}.

\begin{figure}[htbp]
\begin{center}
\includegraphics[width=0.9\linewidth]{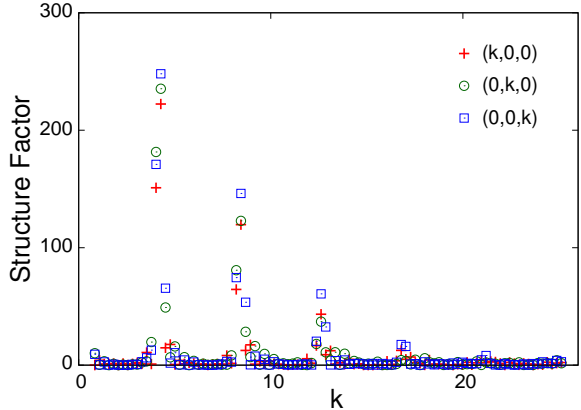}
\end{center}
\caption{(Color online)
Structure factor in the solid phase ($N_i=80$),
computed at $t=8000$,
for $\bm{k} = (k,0,0), (0,k,0), (0,0,k)$.
}
\label{fig3}
\end{figure}

We now study the solid phase by increasing $\rho_e$.
We squeeze $L_y$ and $L_z$ from $20$ to $12$, 
while keeping $L_x=15$.
This yields $\rho_e = 0.621$ ($> \rho_e^{\rm sl}= 0.545$).
In Fig.~\ref{fig3} we plot the structure factor
$
S(\bm{k}) 
=
N_e^{-1} \sum_{j, l =1}^{N_e} e^{- i \bm{k}\cdot (\bm{q}^j - \bm{q}^l)},
$
obtained by the MD simulation, 
where $\bm{q}^j$ 
is the position of the $j$-th $e$ particle.
It clearly shows that $e$ particles are in a solid phase.
We have computed $\Xi'(\omega)$ of this state, 
as shown by squares in Fig.~\ref{fig2}. 
It is seen that $\Xi'(\omega) \to 0$ as $\omega \to 0$, 
i.e., 
the solid is much more rigid than the liquid
at low frequencies.
With increasing $\omega$, on the other hand, 
$\Xi'(\omega)$ of the solid (squares) increases whereas that of the liquid 
(circles) decreases.
As a result, 
$\Xi'(\omega)$ of the solid
becomes much larger than that of the liquid around $\omega \sim 1$,
at which $\Xi'(\omega)$ of the solid has a peak.
That is, 
{\em the solid is much more flexible in this frequency region}.
We have thus confirmed our conclusion
that is derived from the universal properties of $\Xi'(\omega)$. 
Furthermore, 
numerical integration of 
$\int \Xi'(\omega) d \omega/\pi =2 \int \omega \Xi'(\omega) d \ln(\omega)/\pi$
for the solid gives 
$\simeq 0.98$ 
\cite{cutoff}, 
in consistency with the sum rule.
This evidences that even in the solid phase 
our MD simulation well describes NE states driven by 
$\bm{f}(t)$ \cite{SY2010}.

{\em Physics.---}
We have seen that in a solid phase $\Xi'(\omega)$ has a peak 
at a high frequency, which makes the solid more flexible than the liquid.
The peak also keeps the sum value constant, compensating for 
reduction of $\Xi'(\omega)$ at low frequencies.
To explore physics of the peak,
we study its dependence on the configuration and 
the number density $n_i$ of impurities. 
The peak would be insensitive to them,
as long as $n_i \ll n_e = N_e/(L_x L_y L_z)$ \cite{sukima},
if it came purely 
from {\em intrinsic} excitation modes in the solid of $e$ particles.

When the impurity configuration is changed, 
we find that the peak position shifts slightly 
(by $\sim \pm 0.1$ \cite{spm}).
This indicates that 
the peak is not intrinsic to the $e$-particle solid.
To see this more clearly, 
we study the cases of lower $n_i$, keeping $n_e$ constant.
In the above case of solid, we have taken 
$n_i  = 80/2160=0.0370\cdots$ and 
$n_e = 320/2160=0.148\cdots$.
The triangles in Fig.~\ref{fig2} show $\Xi'(\omega)$
for lower $n_i$, where $n_i  = 40/2160=0.0185\cdots$.
Numerical integration of $\Xi'(\omega)$ gives 
$\simeq 0.99$ 
\cite{cutoff}, 
and the sum rule is satisfied also in this case,
supporting correctness of our MD simulation.
It is seen that 
the peak shifts significantly to a lower frequency.
For confirmation, 
we further decrease $n_i$ to $n_i=0$.
Although our measure of rigidity is {\em not} applicable to 
this limiting case,
we can see the physics.
The crosses in Fig.~\ref{fig2} show $\Xi'(\omega)$ for $n_i=0$.
Numerical integration of $\Xi'(\omega)$ gives 
$\simeq 0.95$, 
again supporting correctness of our MD simulation
\cite{cutoff}.
It is seen that 
the peak of $\Xi'(\omega)$ at a high frequency almost disappears.
Hence, we have confirmed that the peak is {\em not} 
purely intrinsic to the solid of $e$-particles.
On the other hand, obviously, it is not 
purely extrinsic, either (because, e.g., it disappears in the liquid phase).
Therefore, 
the peak is determined by many factors, including both intrinsic 
ones (such as phonon modes) and extrinsic ones (such as the density of pinning centers).

Interestingly, 
$\Xi'(0)$ for the solid with $n_i=0$ 
does not vanish, 
and $\Xi'(\omega)$ behaves as if $e$ particles were in 
a liquid phase.
This is understood as due to the 
`sliding motion' of the solid \cite{sliding1,sliding2}.
That is, the solid can slide (i.e, move without deformation) 
in the unrealistic case where
pinning centers are completely absent.
As a result, $\Xi'(0)$ for the solid becomes finite for $n_i = 0$, 
and $\Xi'(\omega)$ behaves like that of a liquid.
When $n_i > 0$, on the other hand, 
such a sliding motion is impossible 
because the solid is pinned by impurities,
and $\Xi'(0)$ vanishes. 
By contrast, a liquid has finite $\Xi'(0)$ even when $n_i > 0$,
as shown by circles of Fig.~\ref{fig2},
because it is flexible at $\omega \simeq 0$ so that it is not pinned.


{\em Discussions.---}
Since our main result holds quite generally, 
it can be tested experimentally in diverse systems,
such as electron systems, cold atoms \cite{ca1,ca2,ca3,ca4},
and colloidal suspensions.
Note that measured $\Xi'(\omega)$ 
often contains contributions 
not only from particles of interest (which 
undergo a phase transition) but also 
from other particles or excitations.
Hence, it is necessary to extract the contribution from 
particles of interest.

One sometimes extracts contributions {\em only from a single band}.
Since the energy scale is strongly restricted in such a case, 
the `restricted sum rule' holds
instead of the `full sum rule' (\ref{sumrule}).
In this case, we take for $I(t)$ 
the current integrated over the position $x$ of the cross section
rather than the `crystal momentum' (times a constant), 
because the latter is not directly related to transport.
In a way similar to derivation of the restricted sum rule for 
the optical 
conductivity \cite{t0,t1,t4,Basov,Vescoli,SY2011}, 
we obtain the one for the longitudinal admittance as \cite{spm}
\begin{equation}
\int_{-\infty}^{\infty} \Xi'(\omega) {d \omega \over \pi}
=
\sum_{\bm{k},\sigma}
m^{-1}(\bm{k}) 
\langle \hat{n}_{\bm{k} \sigma} \rangle.
\label{sumrule_sb}\end{equation}
Here, 
$m^{-1}(\bm{k})$ is a diagonal element of the inverse mass tensor,
and $\langle \hat{n}_{\bm{k} \sigma} \rangle$ is the occupation
of the Bloch state with wavevector $\bm{k}$ and spin $\sigma$.
In such a case, 
only when electrons occupy low-energy states of the band
(in all phases)
the sum value becomes constant,
because 
$
\sum_{\bm{k},\sigma} m^{-1}(\bm{k}) \langle \hat{n}_{\bm{k} \sigma} \rangle
\simeq 
m^{-1}(\bm{0}) \sum_{\bm{k},\sigma}  \langle \hat{n}_{\bm{k} \sigma} \rangle 
=
m^{-1}(\bm{0}) N_e.
$
Under this condition, 
the results of the present paper hold also in such 
a single-band case.

\begin{acknowledgments}
We are grateful to K. Asano, Y. Kato and H. Hayakawa for helpful discussions.
This work was supported by KAKENHI from the Japan Society for the Promotion of Science, Grant Nos.~26287085 and 15H05700.
\end{acknowledgments}


\begin{thebibliography}{99}
\bibitem{interaction}
When interparticle potentials are changed, 
the $\omega$ dependence of $\Xi'(\omega)$
and the liquid-solid transition point change. 
However, the sum value and our main result do not change at all.
\bibitem{cutoff}
The integral value of $\Xi'(\omega)$ is slightly underestimated
because we have computed it by taking 
the upper bound of the integral to $\omega=10$,
above which $\Xi'(\omega)$ is vanishingly small 
(and hence relative errors are large).
\bibitem{sukima}
Recall that an impurity has vanishing radius.
In a solid phase, we distribute impurities in vacant spaces among $e$ particles.
\end{thebibliography}

\end{document}